\documentclass{llncs}
\usepackage{graphicx}
\usepackage{color}
\usepackage{url}

\newcommand{\acetext}[1]{\textsf{\footnotesize#1}}

\begin{document}

\title{
  How Controlled English can Improve\\
  Semantic Wikis
}

\author{
  Tobias Kuhn
}

\institute{
  Department of Informatics \& Institute of Computational Linguistics,\\
  University of Zurich, Switzerland\\
  \texttt{tkuhn@ifi.uzh.ch}\\
  \texttt{http://www.ifi.uzh.ch/cl/tkuhn}
}

\maketitle

\begin{abstract}
The motivation of semantic wikis is to make acquisition, maintenance, and mining of formal knowledge simpler, faster, and more flexible. However, most existing semantic wikis have a very technical interface and are restricted to a relatively low level of expressivity. In this paper, we explain how AceWiki uses controlled English --- concretely Attempto Controlled English (ACE) --- to provide a natural and intuitive interface while supporting a high degree of expressivity. We introduce recent improvements of the AceWiki system and user studies that indicate that AceWiki is usable and useful.
\end{abstract}

\section{Introduction}

We present an approach how semantic wikis and controlled natural language can be brought together. This section gives a short introduction into the fields of semantic wikis and controlled natural languages.

\subsection{Semantic Wikis}

Semantic wikis are a relatively new field of research that started in 2004 when semantic wikis were introduced in \cite{tazzoli:iswc04} describing the PlatypusWiki system. During the last years, an active community emerged and many new semantic wiki systems were presented. Semantic wikis combine the philosophy of wikis (i.e. quick and easy editing of textual content in a collaborative way over the Web) with the concepts and techniques of the Semantic Web (i.e. enriching the data on the Web with well-defined meaning). The idea is to manage formal knowledge representations within a wiki environment.

Generally, two types of semantic wikis can be distinguished: On the one hand, there are text-centered approaches that enrich classical wiki environments with semantic annotations. On the other hand, logic-centered approaches use semantic wikis as a form of online ontology editors. Semantic MediaWiki \cite{kroetzsch:websemantics2007}, IkeWiki \cite{schaffert:stica2006}, SweetWiki \cite{buffa08sweetwiki}, and HyperDEWiki \cite{iswc2008pd_schwabe} are examples of text-centered semantic wikis, whereas OntoWiki \cite{auer:iswc2006} and myOntology \cite{siorpaes07myontology} are two examples of logic-centered semantic wikis. Web-Prot\'eg\'e \cite{tudorache:owled2008} is the Web version of the popular Prot\'eg\'e ontology editor and can be seen as another example of a logic-centered semantic wiki, even though its developers do not call it a ``semantic wiki''. In general, there are many new web applications that do not call themselves ``semantic wikis'' but exhibit many of their characteristic properties. Freebase\footnote{see \cite{bollacker:sigmod08} and \url{http://www.freebase.com}}, Knoodl\footnote{\url{http://knoodl.com}}, SWIRRL\footnote{\url{http://www.swirrl.com}}, and Twine\footnote{\url{http://www.twine.com}} are some examples.

Semantic wikis seem to be a very promising approach to get the domain experts better involved in the creation and maintenance of ontologies. Semantic wikis could increase the number and quality of available ontologies which is an important step into the direction of making the Semantic Web a reality. However, we see two major problems with the existing semantic wikis. First, most of them have a very technical interface that is hard to understand and use for untrained persons, especially for those who have no particular background in formal knowledge representation. Second, existing semantic wikis support only a relatively low degree of expressivity --- mostly just ``subject predicate object''-structures --- and do not allow the users to assert complex axioms. These two shortcomings have to be overcome to enable average domain experts to manage complex ontologies through semantic wiki interfaces.

In this paper, we will argue for using controlled natural language within semantic wikis. The Wiki@nt\footnote{see \cite{bao:eon04} and \url{http://tw.rpi.edu/dev/cnl/}} system follows a similar approach. It uses controlled natural language (concretely Rabbit and ACE) for verbalizing OWL axioms. In contrast to our approach, users cannot create or edit the controlled natural language sentences directly but only the underlying OWL axioms in a common formal notation. Furthermore, no reasoning takes place in Wiki@nt.

\subsection{Attempto Controlled English}

Controlled natural languages are subsets of natural languages that are controlled (in both syntax and semantics) in a way that removes or reduces the ambiguity of the language. Recently, several controlled natural languages have been proposed for the Semantic Web \cite{schwitter:owled2008dc}. The idea is to represent formal statements in a way that looks like natural language in order to make them better understandable to people with no background in formal methods.

Attempto Controlled English (ACE)\footnote{see \cite{fuchs:reasoningweb2008} and \url{http://attempto.ifi.uzh.ch}} is a controlled subset of English. While looking like natural English, it can be translated automatically and unambiguously into logic. Thus, every ACE text has a single and well-defined formal meaning. A subset of ACE has been used as a natural language front-end to OWL with a bidirectional mapping from ACE to OWL \cite{kaljurand:phd}. This mapping covers all of OWL 2 except data properties and some very complex class expressions.

ACE supports a wide range of natural language constructs: nouns (e.g. ``\acetext{country}''), proper names (``\acetext{Zurich}''), verbs (``\acetext{contains}''), adjectives (``\acetext{rich}''), singular and plural noun phrases (``\acetext{a person}'', ``\acetext{some persons}''), active and passive voice (``\acetext{owns}'', ``\acetext{is owned by}''), pronouns (``\acetext{she}'', ``\acetext{who}'', ``\acetext{something}''), relative phrases (``\acetext{who is rich}'', ``\acetext{that is owned by John}''), conjunction and disjunction (``\acetext{and}'', ``\acetext{or}''), existential and universal quantifiers (``\acetext{a}'', ``\acetext{every}''), negation (``\acetext{no}'', ``\acetext{does not}''), cardinality restrictions (``\acetext{at most 3 persons}''), anaphoric references (``\acetext{the country}'', ``\acetext{she}''), questions (``\acetext{what borders Switzerland?}''), and much more. Using these elements, one can state ACE sentences like for example
\begin{quote}
  \acetext{Every person who writes a book is an author.}
\end{quote}
that can be translated into its logical representation:
\begin{quote}
  $\forall A \forall B ( person(A) \wedge write(A, B) \wedge book(B) \rightarrow author(A) )$
\end{quote}
In the functional-style syntax of OWL, the same statement would have to be expressed as follows:
\begin{quote}
\footnotesize
\begin{verbatim}
SubClassOf(
  IntersectionOf(
    Class(:person)
    SomeValuesFrom(
      ObjectProperty(:write)
      Class(:book)
    )
  )
  Class(:author)
)
\end{verbatim}
\end{quote}
This example shows the advantage of controlled natural languages like ACE over other logic languages. While the latter two statements require a considerable learning effort to be understood, the statement in ACE is very easy to grasp even for a completely untrained reader. We could show in an experiment that untrained users (who have no particular background in knowledge representation or computational linguistics) are able to understand ACE sentences very well and within very short time \cite{kuhn:cnl2009}.

\section{AceWiki}

\begin{figure}[tb]
  \begin{center}
    \includegraphics[width=\textwidth]{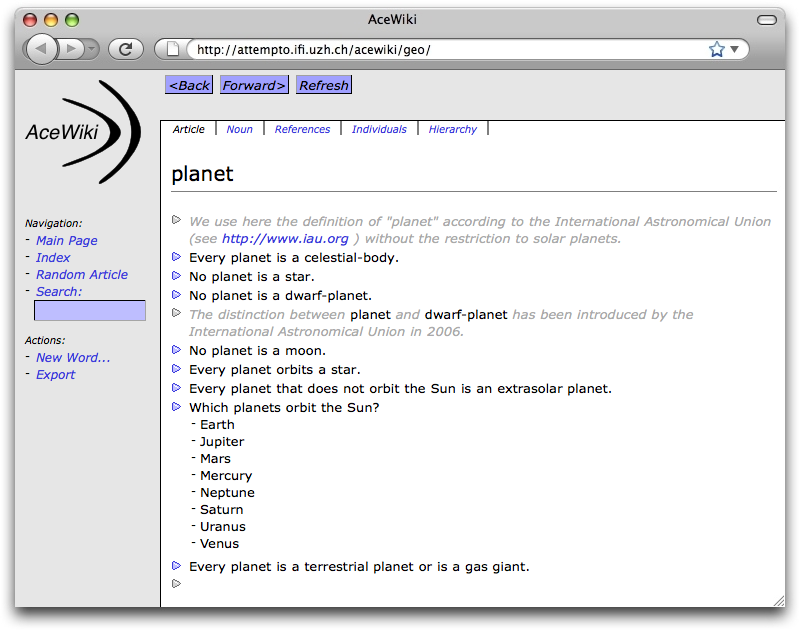}
    \caption{This screenshot of the AceWiki interface shows an article about planets. Articles in AceWiki consist of declarative ACE sentences and ACE questions (both in black color) and of unrestricted natural language comments (in gray color).}
    \label{fig:window}
  \end{center}
\end{figure}

We developed AceWiki that is a logic-centered semantic wiki that tries to solve the identified problems of existing semantic wikis by using a subset of ACE as its knowledge representation language. The goal of AceWiki is to show that semantic wikis can be more natural and at the same time more expressive than existing systems. Figure \ref{fig:window} shows a screenshot of the AceWiki interface. The general approach is to provide a simple and natural interface that hides all technical details.

AceWiki has been introduced in \cite{kuhn:swui2008} and \cite{kuhn:semwiki2008}. Since then, many new features have been implemented: support for transitive adjectives, abbreviations for proper names, passive voice for transitive verbs, support for comments, client-side OWL export, a completely redesigned lexical editor, and proper persistent storage of the wiki data. There is a public demo available\footnote{see \url{http://attempto.ifi.uzh.ch/acewiki}} and the source code of AceWiki can be downloaded under an open source license. However, AceWiki has not yet reached the stage where it could be used for real-world applications. Crucial (but scientifically not so interesting) parts are missing: history/undo facility, user management, and ontology import.

One of the most interesting new features in AceWiki is the support for comments in unrestricted natural language, as it can be seen in Figure \ref{fig:window}. Since it is unrealistic that all available information about a certain topic can be represented in a formal way, such comments can complement the formal ACE sentences. The comments can contain internal and external links, much like the text in traditional non-semantic wikis.

\begin{figure}[tb]
  \begin{center}
    \includegraphics[width=\textwidth]{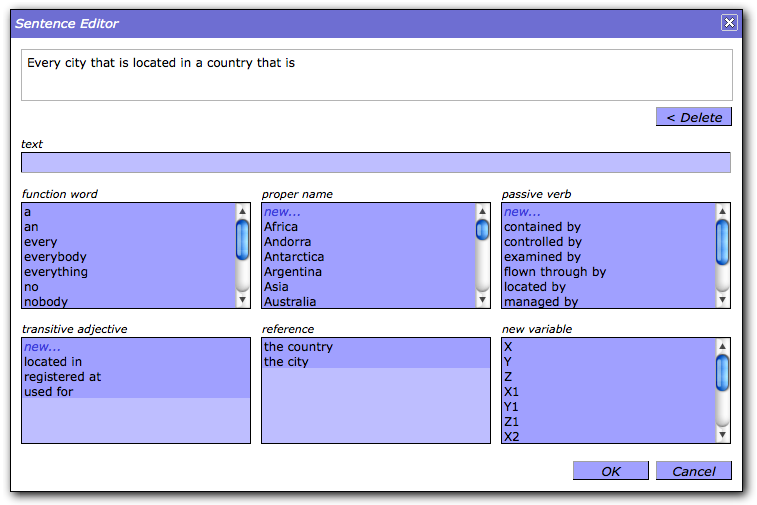}
    \caption{This is the predictive editor of AceWiki. The partial sentence ``\acetext{every city that is located in a country that is ...}'' has already been entered and now the possible continuations are shown. In this way, the users can conveniently create syntactically correct sentences without learning the language in advance.}
    \label{fig:editor}
  \end{center}
\end{figure}

In order to enable the easy creation of ACE sentences, users are supported by an intelligent predictive text editor \cite{alta2008_kuhnschwitter} that is able to look ahead and to show the possible words and phrases to continue the sentence. Figure \ref{fig:editor} shows a screenshot of this editor.

AceWiki supports an expressive subset of ACE. Some examples of sentences that can be created in AceWiki are shown here:
\begin{quote}
  \includegraphics[width=10cm]{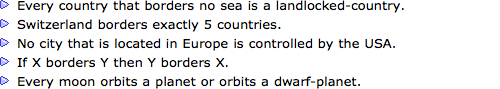}
\end{quote}

AceWiki is designed to seamlessly integrate a reasoner that can give feedback to the users, ensures the consistency of the ontology, can show its semantic structure, and answers queries. At the moment, we are using the OWL reasoner Pellet\footnote{\url{http://clarkparsia.com/pellet/}} and apply the ACE-to-OWL translator that is described in \cite{kaljurand:phd}. However, AceWiki is not restricted to OWL and another reasoner or rule engine might be used in the future.

The subset of ACE that is used in AceWiki is more expressive than OWL, and thus the users can assert statements that have no OWL representation. Because we are using an OWL reasoner at the moment, such statements are not considered for reasoning. In order to make this clear to the users, the sentences that are outside of OWL are marked by a red triangle:
\begin{quote}
  \includegraphics[width=10cm]{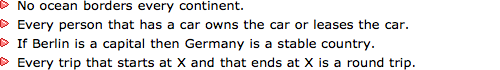}
\end{quote}

The most important task of the reasoner is to check consistency because only consistent ontologies enable to calculate logical entailments. In previous work \cite{kuhn:semwiki2008}, we explain how consistency is ensured in AceWiki by incrementally checking every new sentence that is added.

Not only asserted but also inferred knowledge can be represented in ACE. At the moment, AceWiki shows inferred class hierarchies and class memberships. The hierarchy for the noun ``\acetext{country}'', for example, could look as follows:
\begin{quote}
  \includegraphics[width=10cm]{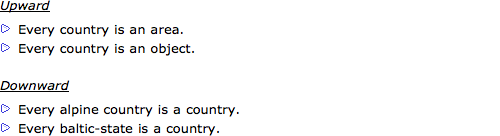}
\end{quote}
Furthermore, ACE questions can be formulated within the articles. Such questions are evaluated by the reasoner and the results are listed directly after the question:
\begin{quote}
  \includegraphics[width=10cm]{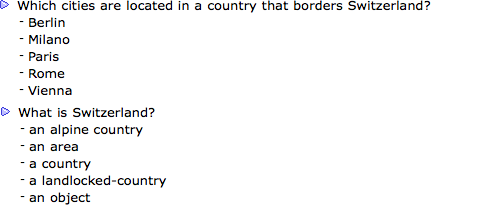}
\end{quote}
If the question asks for a certain individual (represented in ACE by proper names) then the named classes (represented by nouns) of the individual are shown as the answer. In the cases where the question asks for a class (represented by a noun phrase), the individuals that belong to this class are shown as the answer.

Thus, AceWiki uses ACE in different ways: as an expressive knowledge representation language for asserted knowledge, to display entailed knowledge generated by the reasoner, and as a query language.

In AceWiki, words have to be defined before they can be used. At the moment, five types of words are supported: proper names, nouns, transitive verbs, \emph{of}-constructs (i.e. nouns that have to be used with \emph{of}-phrases), and transitive adjectives (i.e. adjectives that require an object). A new feature of AceWiki is that proper names can have an abbreviation that has exactly the same meaning as the long proper name. This is very helpful for proper names that are too long to be spelled out each time.

\begin{figure}[tb]
  \begin{center}
    \includegraphics[width=\textwidth]{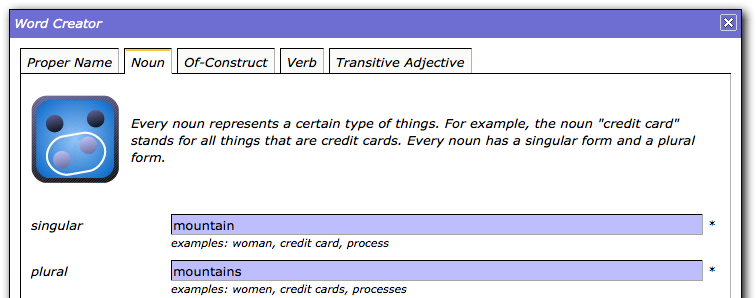}
    \caption{The lexical editor of AceWiki helps the users to define the word forms. The example shows how a new noun ``\acetext{mountain}'' is created.}
    \label{fig:lexicaleditor}
  \end{center}
\end{figure}

Figure \ref{fig:lexicaleditor} shows the lexical editor of AceWiki that helps the users in creating and modifying word forms in an appropriate way. An icon and an explanation in natural language help the users to choose the right category.

\section{Evaluation}

In order to find out how usable and how useful AceWiki is, we performed several tests. From time to time, we set up small usability experiments to test how well normal users are able to cope with the current version of AceWiki. Two such experiments have been performed so far. In order to find out whether AceWiki can be useful in the real world, we additionally conducted a small case study in which we tried to formalize the content of the existing Attempto project website in AceWiki. The results are explained in the following sections. Table \ref{tab:expoverview} shows an overview.
\begin{table}[tb]
\begin{center}
\caption{This table compares the settings of the three tests that have been performed on AceWiki.}
\label{tab:expoverview}
\begin{tabular}{@{\hspace{3pt}}p{3.8cm}@{\hspace{3pt}}|@{\hspace{3pt}}r@{\hspace{3pt}}|@{\hspace{3pt}}r@{\hspace{3pt}}|@{\hspace{3pt}}r@{\hspace{3pt}}|}
 & Experiment 1 & Experiment 2 & Case Study \\
\hline
time & Nov 2007 & Nov 2008 & Nov/Dec 2008 \\
\hline
AceWiki version & 0.2.1 & 0.2.9 & 0.2.10 \\
\hline
number of subjects ($n$) & 20 & 6 & 1 \\
\hline
subjects & mostly students & students & AceWiki developer \\
\hline
level of preexisting knowledge about AceWiki & none & low & highest \\
\hline
domain to be represented & ``the real world'' & universities & Attempto project \\
\hline
\end{tabular}
\end{center}
\end{table}

\subsection{Usability Experiments}

Two usability experiments have been performed so far on AceWiki. The first one took place in November 2007 and has been described and analyzed in \cite{kuhn:swui2008}. The second experiment --- that is introduced here --- was conducted one year later in November 2008. Both experiments have the nature of cheap ad hoc experiments with the goal to get some feedback about possible weak points of AceWiki. Since the settings of the two experiments were different and since the number of subjects was relatively low, we cannot draw strong statistical conclusions. Nevertheless, these experiments can give us valuable feedback about the usability of AceWiki.

In both experiments, the subjects were told to create a formal knowledge base in a collaborative way using AceWiki. The task was just to add correct and meaningful knowledge about the given domain without any further constraints on the kind of knowledge to be added. The subjects --- mostly students --- had no particular background in formal knowledge representation. The domain to be represented was the real world in general in the first experiment, and the domain of universities (i.e. students, departments, professors, etc.) in the second experiment.

In the first experiment, the subjects received no instructions at all how AceWiki has to be used. In the second experiment, they attended a 45 minutes lesson about AceWiki. Another important difference is that the first experiment used an older version of AceWiki where templates could be used for the creation of certain types of sentences (e.g. class hierarchies). This has been removed in later versions because of its lack of generality.

Table \ref{tab:expresults} shows the results of the two experiments. Since the subjects worked together on the same knowledge base and could change or remove the contributions of others, we can look at the results from two perspectives: On the one hand, there is the community perspective where we consider only the final result, not counting the sentences that have been removed at some point and only looking at the final versions of the sentences. On the other hand, from the individuals perspective we count also the sentences that have been changed or removed by another subject. The different versions of a changed sentence count for each of the respective subjects. However, sentences created and then removed by the same subject are not counted, and only the last version counts for sentences that have been changed by the same subject.
\begin{table}[tb]
\begin{center}
\caption{This table shows the results of the first (Exp. 1) and the second (Exp. 2) experiment. The results can be seen from the individuals perspective (ind.) or the community perspective (comm.)}
\label{tab:expresults}
\begin{tabular}{@{\hspace{3pt}}l@{\hspace{3pt}}|@{\hspace{3pt}}l@{\hspace{3pt}}|@{\hspace{3pt}}r@{\hspace{3pt}}r@{\hspace{3pt}}|@{\hspace{3pt}}r@{\hspace{3pt}}r@{\hspace{3pt}}|}
 & & & Exp. 1 & & Exp. 2 \\
 & & ind. & comm. & ind. & comm. \\
\hline
total sentences created & $S$ & 186 & 179 & 113 & 93 \\
\hspace{2mm}correct sentences & $S^+$ & 148 & 145 & 76 & 73 \\
\hspace{4mm}correct sentences that are complex & $S^+_x$ & 91 & 89 & 54 & 51 \\
\hspace{2mm}sentences using ``a'' instead of ``every'' & $S^e$ & 9 & 9 & 23 & 12 \\
\hspace{2mm}sentences using misclassified words & $S^w$ & 9 & 8 & 0 & 0 \\
\hspace{2mm}other incorrect sentences & $S^-$ & 20 & 17 & 14 & 8 \\
\% of correct sentences & $S^+/S$ & 80\% & 81\% & 67\% & 78\% \\
\% of (almost) correct sentences & $(S^+ + S^e)/S$ & 84\% & 86\% & 88\% & 91\% \\
\% of complex sentences & $S^+_x/S^+$ & 61\% & 61\% & 71\% & 70\% \\
\hline
total words created & $w$ & 170 & 170 & 53 & 50 \\
\hspace{2mm}individuals (i.e. proper names) & $w_p$ & 44 & 44 & 11 & 10 \\
\hspace{2mm}classes (i.e. nouns) & $w_n$ & 81 & 81 & 14 & 14 \\
\hspace{2mm}relations total & $w_r$ & 45 & 45 & 28 & 26 \\
\hspace{4mm}transitive verbs & $w_v$ & 39 & 39 & 20 & 18 \\
\hspace{4mm}\emph{of}-constructs & $w_o$ & 6 & 6 & 2 & 2 \\
\hspace{4mm}transitive adjectives & $w_a$ & -- & -- & 6 & 6 \\
sentences per word & $S/w$ & 1.09 & 1.05 & 2.13 & 1.86 \\
correct sentences per word & $S^+/w$ & 0.87 & 0.85 & 1.43 & 1.46 \\
\hline
total time spent (in minutes) & $t$ & 930.9 & 930.9 & 360.2 & 360.2 \\
av. time per subject & $t/n$ & 46.5 & 46.5 & 60.0 & 60.0 \\
av. time per correct sentence & $t/S^+$ & 6.3 & 6.4 & 4.7 & 4.9 \\
av. time per (almost) correct sentence & $t/(S^+ + S^e)$ & 5.9 & 6.0 & 3.6 & 4.2 \\
\hline
\end{tabular}
\end{center}
\end{table}

The first part of the table shows the number and type of sentences the subjects created. In total, the resulting knowledge bases contained 179 and 93 sentences, respectively. We checked these sentences manually for correctness. $S^+$ stands for the number of sentences that are (1) logically correct and (2) sensible to state.

The first criterion is simple: In order to be classified as correct, the sentence has to represent a correct statement about the real world using the common interpretations of the words and applying the interpretation rules of ACE.

The second criterion can be explained best on the basis of the sentences of the type $S^e$. These sentence start with ``\acetext{a} ...'' like for example ``\acetext{a student studies at a university}''. This sentence is interpreted in ACE as having only existential quantification: ``\acetext{there is a student that studies at a university}''. This is certainly a logically correct statement about the real world, but the writer probably wanted to say ``\acetext{every student studies at a university}'' which is a more precise and more sensible statement. For this reason, such statements are not considered correct, even though they are correct from a purely logical point of view.

Sentences of the type $S^e$ have been identified in \cite{kuhn:swui2008} as one of two frequent error types when using AceWiki. The other one --- denoted by $S^w$ --- are sentences using words in the wrong word category like for example ``\acetext{every London is a city}'' where ``\acetext{London}'' has been added as a noun instead of a proper name.

It is interesting that the incorrect sentences of the types $S^e$ and $S^w$ had the same frequency in the first experiment, but evolved in different directions in the second experiment. There was not a single case of $S^w$-mistakes in the second experiment. This might be due to the fact that we learned from the results of the first experiment and enriched the lexical editor with icons and explanations (see Figure \ref{fig:lexicaleditor}).

\begin{figure}[tb]
\begin{center}
\includegraphics[width=\textwidth]{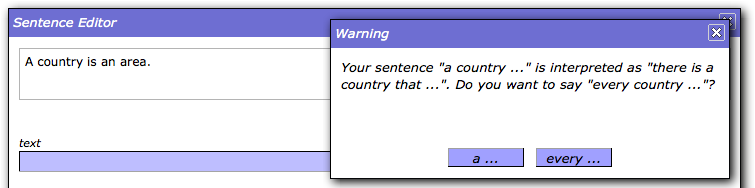}
\caption{This figure shows a solution to the problem that the users often state sentences starting with ``\acetext{a} ...'' when they should say ``\acetext{every} ...''. This is included in the latest version of AceWiki.}
\label{fig:a-to-every}
\end{center}
\end{figure}

On the other hand, the number of $S^e$-mistakes increased. This might be caused by the removal of the templates feature from AceWiki. In the first experiment, the subjects were encouraged to say ``\acetext{every} ...'' because there were templates for such sentences. In the second experiment, those templates were not available anymore and the subjects were tempted to say ``\acetext{a}'' instead of ``\acetext{every}''. This is bad news for AceWiki, but the good news is that there are two indications that we are on the right track nevertheless. First, while none of the $S^e$-sentences has been corrected in the first experiment, almost half of them have been removed or changed by the community during the second experiment. This indicates that some subjects of the second experiment recognized the problem and tried to resolve it. Second, the $S^e$-sentences can be detected and resolved in a very easy way. Almost every sentence starting with ``\acetext{a} ...'' is an $S^e$-sentence and can be corrected just by replacing the initial ``\acetext{a}'' by ``\acetext{every}''. After the second experiment, we added a new feature to AceWiki that asks the users each time they create a sentence of the form ``\acetext{a} ...'' whether it should be ``\acetext{every} ...''. The users can then say whether they really mean ``\acetext{a} ...'' or whether it should be rather ``\acetext{every} ...''. In the latter case the sentence is automatically corrected. Figure \ref{fig:a-to-every} shows a screenshot of the dialog shown to the users. Future experiments will show whether this solves the problem.

An interesting figure is of course the ratio of correct sentences $S^+/S$. As it turns out, the first experiment exhibits the better ratio for both perspectives: 80\% versus 67\% for the individuals and 81\% versus 78\% for the community. However, since $S^e$-sentences are easily detectable and correctable (and hopefully a solved problem with the latest version of AceWiki), it makes sense to have a look at the ratio of ``(almost) correct'' sentences consisting of the correct sentences $S+$ and the $S^e$-sentences. This ratio was better in the second experiment: 84\% versus 88\% for the individuals perspective; 86\% versus 91\% for the community perspective. However, the settings of the experiments do not allow us to draw any statistical conclusions from these numbers. Nevertheless, these results give us the impression that a ratio of correct and sensible statements of 90\% and above is achievable with our approach.

Another important aspect is the complexity of the created sentences. Of course, syntactically and semantically complex statements are harder to construct than simple ones. For this reason, we classified the correct sentences according to their complexity. $S^+_c$ stands for all correct sentences that are complex in the sense that they contain a negation (``\acetext{no}'', ``\acetext{does not}'', etc.), an implication (``\acetext{every}'', ``\acetext{if} ... \acetext{then}'', etc.), a disjunction (``\acetext{or}''), a cardinality restriction (``\acetext{at most 3}'', etc.), or several of these elements. While the ratio of complex sentences was already very high in the first experiment (around 60\%), it was was even higher in the second experiment reaching 70\%. Looking at the concrete sentences the subjects created during the second experiment, one can see that they managed to create a broad variety of complex sentences. Some examples are shown here:
\begin{itemize}
\item \acetext{Every lecture is attended by at least 3 students.}
\item \acetext{Every lecturer is a professor or is an assistant.}
\item \acetext{Every professor is employed by a university.}
\item \acetext{If X contains Y then X is larger\_than Y.}
\item \acetext{If somebody X likes Y then X does not hate Y.}
\item \acetext{If X is a student and a professor knows X then the professor hates X or likes X or is indifferent\_to X.}
\end{itemize}
The last example is even too complex to be represented in OWL. Thus, the AceWiki user interface seems to scale very well in respect to the complexity of the ontology.

The second part of Table \ref{tab:expresults} shows the number and types of the words that have been created during the experiment. All types of words have been used by the subjects with the exception that transitive adjectives were not supported by the AceWiki version used for the first experiment. It is interesting to see that the first experiment resulted in an ontology consisting of more words than correct sentences, whereas in the second experiment the number of correct sentences clearly exceeds the number of words. This is an indication that the terms in the second experiment have been reused more often and were more interconnected.

The third part of Table \ref{tab:expresults} takes the time dimension into account. On average, each subject of the first experiment spent 47 minutes, and each subject of the second experiment spent 60 minutes. The average time per correct sentence that was around 6.4 minutes in the first experiment was much better in the second experiment being only 4.9 minutes. We consider these time values very good results, given that the subjects were not trained and had no particular background in formal knowledge representation.

In general, the results of the two experiments indicate that AceWiki enables unexperienced users to create complex ontologies within a short amount of time.

\subsection{Case Study}

The two experiments presented above seem to confirm that AceWiki can be used easily by untrained persons. However, usability does not imply the usefulness for a particular purpose. For that reason, we performed a small case study to exemplify how an experienced user can represent a strictly defined part of real-world knowledge in AceWiki in a useful way.

The case study presented here consists of the formalization of the content of the Attempto website\footnote{\url{http://attempto.ifi.uzh.ch}} in AceWiki. This website contains information about the Attempto project and its members, collaborators, documents, tools, languages, and publications, and the relations among these entities. Thus, the information provided by the Attempto website is a piece of relevant real-world knowledge.

In the case study to be presented, one person --- the author of this paper who is the developer of AceWiki --- used a plain AceWiki instance and filled it with the information found on the public Attempto website. The goal was to represent as much as possible of the original information in a natural and adequate way. This was done manually using the AceWiki editor without any kind of automation.

Table \ref{tab:casestudyresults} shows the results of the case study. The formalization of the website content took less than six hours and resulted in 538 sentences. This gives an average time per sentence of less than 40 seconds. These results give us some indication that AceWiki is not only usable for novice users but can also be used in an efficient way by experienced users.

Most of the created words are proper names (i.e. individuals) which is not surprising for the formalization of a project website. The ratio of complex sentences is much lower than the ones encountered in the experiments but with almost 20\% still on a considerable level.
\begin{table}[tb]
\begin{center}
\caption{This table lists the results of the AceWiki case study where the content of the Attempto website was formalized in AceWiki.}
\label{tab:casestudyresults}
\begin{tabular}{@{\hspace{3pt}}l@{\hspace{3pt}}|@{\hspace{3pt}}l@{\hspace{3pt}}|@{\hspace{3pt}}r@{\hspace{3pt}}@{\hspace{3pt}}l@{\hspace{3pt}}|}
\hline
total sentences created & $S$ & 538 & \\
\hspace{2mm}complex sentences & $S_x$ & 107 & \\
\% of complex sentences & $S_x/S$ & 19.9\% & \\
\hline
total words created & $w$ & 261 & \\
\hspace{2mm}individuals (i.e. proper names) & $w_p$ & 184 & \\
\hspace{2mm}classes (i.e. nouns) & $w_n$ & 46 & \\
\hspace{2mm}relations total & $w_r$ & 31 & \\
\hspace{4mm}transitive verbs & $w_v$ & 11 & \\
\hspace{4mm}\emph{of}-constructs & $w_o$ & 13 & \\
\hspace{4mm}transitive adjectives & $w_a$ & 7 & \\
sentences per word & $S/w$ & 2.061 & \\
\hline
time spent (in minutes) & $t$ & 347.8 & (= 5.8 h) \\
av. time per sentence & $t/S$ & 0.65 & (= 38.8 s) \\
\hline
\end{tabular}
\end{center}
\end{table}

\begin{figure}[tb]
\begin{center}
\includegraphics[width=\textwidth]{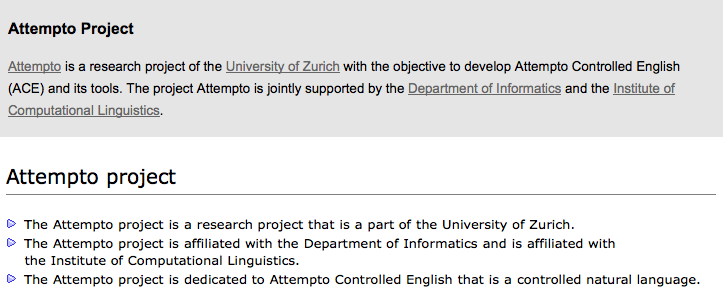}
\caption{This figure shows an example text that occurs on the Attempto website (top) and how it was represented in AceWiki (bottom).}
\label{fig:casestudycomp}
\end{center}
\end{figure}

Basically, all relevant content of the Attempto website could be represented in AceWiki. Of course, the text could not be taken over verbatim but had to be rephrased. Figure \ref{fig:casestudycomp} exemplary shows how the content of the website was formalized. The resulting ACE sentences are natural and understandable.

However, some minor problems were encountered. Data types like strings, numbers, and dates would have been helpful but are not supported. ACE itself has support for strings and numbers, but AceWiki does not use this feature so far. Another problem was that the words in AceWiki can consist only of letters, numbers, hyphens, and blank spaces\footnote{Blank spaces are represented internally as underscores.}. Some things like publication titles or package names contain colons or dots which had to be replaced by hyphens in the AceWiki representation. We plan to solve these problems by adding support for data types and being more flexible in respect to user-defined words.

\begin{figure}[tb]
\begin{center}
\includegraphics[width=\textwidth]{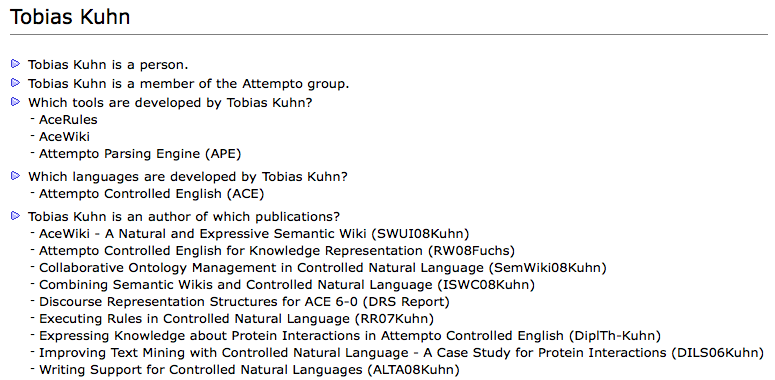}
\caption{This figure shows an exemplary wiki article that resulted from the case study. Inline queries are used to generate content that is automatically updated.}
\label{fig:casestudyex}
\end{center}
\end{figure}

Figure \ref{fig:casestudyex} shows a wiki article that resulted from the case study. It shows how inline queries can be used for automatically generated and updated content. This is an important advantage of such semantic wiki systems. The knowledge has to be asserted once but can be displayed at different places. In the case of AceWiki, such automatically created content is well separated from asserted content in a natural and simple manner by using ACE questions.

As can be seen on Figure \ref{fig:casestudyex}, the abbreviation feature for proper names has been used extensively. The answer lists show the abbreviations in parentheses after the long proper names. The most natural name for a publication, for example, is its title. However, sentences that contain a spelled-out publication title become very hard to read. In such cases, abbreviations are defined which can be used conveniently to refer to the publication.

The fact that the AceWiki developer is able to use AceWiki in an efficient way for representing real world knowledge does of course not imply that every experienced user is able to do so. However, we can see the results as an upper boundary of what is possible to achieve with AceWiki, and the results show that AceWiki \emph{in principle} can be used in an effective way.

\section{Conclusions}

We presented the AceWiki system that should solve the problems that existing semantic wikis do not support expressive ontology languages and are hard to understand for untrained persons. AceWiki shows how semantic wikis can serve as online ontology editors for domain experts with no background in formal methods.

The two user experiments indicate that unexperienced users are able to deal with AceWiki. The subjects managed to create many correct and complex statements within a short period of time. The presented case study indicates that AceWiki is suitable for formalization tasks of the real world and that it can be used --- in principle --- by experienced users in an efficient way. Still, more user studies are needed in the future to prove our claim that controlled natural language improves the usability of semantic wikis.

In general, we showed how controlled natural language can bring the Semantic Web closer to the end users. The full power of the Semantic Web can only be exploited if a large part of the Web users are able to understand and extend the semantic data.

\bibliography{attempto}
\bibliographystyle{plain}

\end{document}